\documentclass[journal,twocolumn]{IEEEtran}
\usepackage{amssymb}
\usepackage{graphicx}
\usepackage{multirow}
\usepackage{hyperref}
\usepackage{graphics} 
\usepackage{tabularx,booktabs}
\usepackage{times} 
\usepackage{pbox}
\usepackage{rotating}
\usepackage[table, dvipsnames]{xcolor}
\usepackage{xcolor}
\usepackage{array}
\usepackage{makecell}
\usepackage{booktabs}

\usepackage{colortbl}
\usepackage{pdflscape}
\usepackage{afterpage}
\usepackage{xcolor,colortbl}

\definecolor{DarkGray}{gray}{0.80}
\definecolor{LightGray}{gray}{0.95}
\definecolor{LightCyan}{rgb}{0.88,1,1}
\usepackage[dvipsnames]{xcolor}
\definecolor{mypink2}{RGB}{219, 48, 122}

\usepackage[font=small,labelfont=bf,
   justification=justified,
   format=plain]{caption}
\usepackage{amssymb}
\usepackage{pifont}
\usepackage[utf8x]{inputenc}
\usepackage{url}
\usepackage{balance}

\usepackage[inline]{enumitem}

\newif\ifremarks

\remarksfalse 
\remarkstrue 

 \date{}

\balance
\begin{document}
\title{A Security Risk Assessment Method for  Distributed Ledger Technology (DLT) based Applications: Three Industry Case Studies }
\author{Elena Baninemeh, Marre Slikker, Katsiaryna Labunets, Slinger Jansen \thanks{E. Baninemeh, M. Slikker, S. Jansen, and K. Labunets are with the Department of Information and Computer Science at Utrecht University, Utrecht, the Netherlands (e-mail:\{e.baninemeh, m.slikker, k.labunets, slinger.jansen)}}

\maketitle
\balance

\begin{abstract}
Distributed ledger technologies have gained significant attention and adoption in recent years. Despite various security features distributed ledger technology provides, they are vulnerable to different and new malicious attacks, such as selfish mining and Sybil attacks. While such vulnerabilities have been investigated, detecting and discovering appropriate countermeasures still need to be reported. Cybersecurity knowledge is limited and fragmented in this domain, while distributed ledger technology usage grows daily. Thus, research focusing on overcoming potential attacks on distributed ledgers is required. This study aims to raise awareness of the cybersecurity of distributed ledger technology by designing a security risk assessment method for distributed ledger technology applications. We have developed a database with possible security threats and known attacks on distributed ledger technologies to accompany the method, including sets of countermeasures. We employed a semi-systematic literature review combined with method engineering to develop a method that organizations can use to assess their cybersecurity risk for distributed ledger applications. The method has subsequently been evaluated in three case studies, which show that the method helps to effectively conduct security risk assessments for distributed ledger applications in these organizations.


\end{abstract}
\IEEEpeerreviewmaketitle
\section{Introduction}

Blockchain technology has received significant attention recently, as it offers a reliable decentralized infrastructure for all kinds of business transactions~\cite{farshidi2020multi}. Distributed Ledger Technology (DLT) can provide an open, decentralized, fault-tolerant transaction mechanism. Blockchain, for instance, is a distributed ledger technology that has attracted considerable attention from both industry and academia since it was initially introduced for Bitcoin~\cite{nakamoto2008peer} to support the exchange of cryptocurrency~\cite{wan2021smart}. DLT is disrupting today's industry at a breakneck pace with the potential to change the world~\cite{ogiela2018security} and offers excellent potential to decentralize operations in collaborative business networks and may even enable new business models~\cite{schaffers2018relevance}.

DLT has been widespread beyond cryptocurrencies or financial usage. The applications extend to several domains, including integrity verification, governance, IoT, healthcare management, privacy and security, and supply chain management~\cite{benassy2021eventually}. 
DLT-based applications refer to software solutions and platforms that utilize a decentralized, distributed, and tamper-resistant ledger to record and manage data, transactions, or assets across a network of multiple nodes or participants in the mentioned domains\cite{natarajan2017distributed}.
Benefits of DLT include cost reduction and increased transparency in information sharing between organizations~\cite{chia2018rethinking,vo2018internet}. Many research papers have pointed out the security and trust aspects of DLT, such as~\cite{hou2021trustseco},~\cite{gojka2021security},~\cite{ogiela2018security}. Nowadays, the security posture of  Blockchain and DLT remains one of the critical topics in the industry and distributed services~\cite{ogiela2018security}. DLT offers extensive opportunities such as financial services~\cite{maull2017distributed}; however, organizations that use DLT-based applications face challenges and limitations such as security and privacy compliance and governance issues that still haven't been thoroughly investigated and addressed~\cite{pelt2021defining}.
Vulnerabilities and weaknesses lead to the execution of various security threats to the distributed ledger platforms; while such vulnerabilities have been thoroughly investigated in extant research, understanding, and development of appropriate countermeasures are still in their infancy. Thus, an examination of the security of DLT-based applications can be valuable~\cite{campbell2020need}.

The literature lacks a security risk assessment method for DLT-based applications. It is essential to design such a method to support the software development lifecycle of such applications to enable creators of DLT-based applications to rapidly assess the risks that are associated with their technology and introduce countermeasures to prevent them. 

DLTs offer various security features such as high availability, fault tolerance, and tamper resistance to protect transaction data. However, DLTs are still vulnerable to cyberattacks, and their maturity level is insufficient to ensure the necessary security for business applications. Therefore, additional conditions and requirements must be considered when applying DLTs in a business context to enhance their security and protect against cyber threats.

After conducting a literature review, it was found that most studies on Blockchain provide an overview of attacks. However, studies that focus on security in DLTs only define a limited set of attacks on DLTs. As a result, definitions of attacks and potential risks in DLTs are scattered across the literature, and most studies do not provide a countermeasure for each attack. Thus, developing a knowledge base that collects potential attacks for DLTs and suggests corresponding countermeasures to enhance DLT security is crucial.

In this study, we have developed a comprehensive knowledge base including 115 attacks on DLTs and identified the specific assets that are at risk for each attack. Our objective is to enhance understanding and awareness of the cybersecurity challenges in the DLT domain. We conducted a literature review to achieve this goal and designed a novel risk assessment method specifically for DLT-based applications. We further refined the method by performing method engineering and then evaluated its effectiveness through three case studies.

The structure of this article is as follows: Section~\ref{BACKGROUND} defines DLT and describes DLT abstraction layers. Section~\ref{RESEARCHMETHOD} formulates the research objective, defines the study's research questions, and explains our research method. Section~\ref{RELATEDWORK} positions the proposed method in this study among the other risk assessment method in the literature. The rest of the paper reports on the following:  Section~\ref{SRAMDA} present a risk assessment method for DLT-based application. Section~\ref{CaseStudies} explains our empirical observations in the context of three real-world case studies that have been conducted to evaluate our proposed method. Section~\ref{Analysis} analyzes the result of case studies and answer research questions. Section~\ref{Discussion} discusses the limitations of the study. The conclusions are formed ~\ref{CONCLUSION}


\section{Background}\label{BACKGROUND}

A  distributed ledger is a secure data structure that resides across multiple computing devices,  spread across locations or regions with no centralized controller. This is a new way of keeping track of who owns financial, physical, or electronic assets~\cite{yaga2019blockchain}.

\begin{figure}[htbp]
\centering
\includegraphics[width=0.47\textwidth]{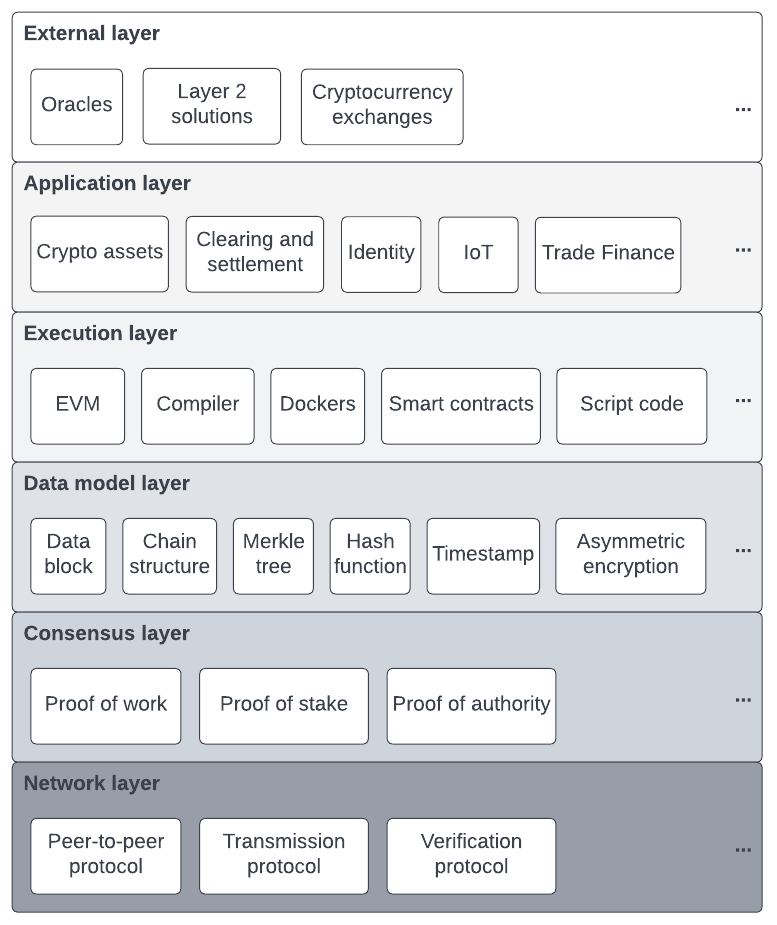}
\caption[DLTstructure]{DLT abstraction layers with example components per layer \cite{TelecommunicationUnion}, \cite{gourisetti2021standardization}}
\label{fig:DLTstructure}
\end{figure}

A DLT consists of six layers: (i) the network layer, (ii) the consensus layer, (iii) the data model layer, (iv) the execution layer, (v) the application layer, and (vi) the external layer~\cite{croman2016scaling}. Figure~\ref{fig:DLTstructure} shows the DLT layers.

\noindent\textbf{Network layer:} The network layer corresponds to the communication infrastructure that is needed to facilitate transactions, information, and data sharing between the nodes. Protocols and methods to facilitate discovery and communication between the peer nodes belong in this layer. If nodes are expected to transact by digitally signing data-in-transit or engage in verification and validation of the transactions, such processes should be defined in this layer~\cite{gourisetti2021standardization}. Risks in this layer are centered around the infrastructure and connectivity of the DLT network. They include threats such as denial-of-service attacks, routing attacks, and network congestion that can disrupt data transmission and compromise the integrity of the network \cite{homoliak2020security}. 

\noindent\textbf{Consensus layer:} The consensus layer is a critical component of DLT technologies, especially blockchains. This layer facilitates distributed trust, ownership, and control. There are two common consensus types: (1) proof-based consensus and (2) voting-based consensus. Consensus has two main properties: (1) indicates an agreement among the distributed nodes and synchronizes them, and (2) validates transactions and ensures reliable and fault-tolerant operations~\cite{rauchs2018distributed}. Risks at this layer pertain to the consensus mechanism that validates transactions and maintains the integrity of the ledger. They include 51\% attacks, protocol deviations, and the potential for forks, which can undermine the trust and reliability of the consensus process, leading to data manipulation and inconsistencies within the ledger \cite{homoliak2019security}.

\noindent\textbf{Data model layer:} This layer handles functions and operations related to the blockchain creation itself in addition to ledger maintenance tasks. Note that this layer does not define the final ledger state. A global consensus is required to approve the final transactions and block creations. However, the process of grouping the transactions into the block, creating a block (or appending to the ledger), and maintaining a common state of the ledger are handled in this layer~\cite{gourisetti2021standardization}. Risks at the Data model layer relate to the structure and management of data within the DLT. They include data integrity risks, data privacy risks, and data access risks, which can agree to the accuracy, confidentiality, and security of the data stored within the ledger \cite{putz2020detecting}.

\noindent\textbf{Execution layer:} This layer contains the DLT rules and program logic such as smart contracts, chain code, etc. The software applications from the application layer trigger the code and rules in the execution layer and instruct the code in the execution layer, resulting in the execution of a transaction. In cases where the execution layer code requires data from off-chain databases, the code can trigger oracles that reside in the application layer (or between the application layer and the outside world) to fetch data/information from off-chain sources to the execution layer code~\cite{rauchs2018distributed}. Risks at this layer involve the execution of transactions and smart contracts within the DLT. They include privacy breaches, smart contract vulnerabilities, and transaction errors, which can result in the exposure of sensitive data, financial losses, and the exploitation of coding flaws, impacting the reliability and security of the DLT system \cite{putz2020detecting}. 

\noindent\textbf{Application layer:} This layer contains applications, software, scripts, and programs that the users can use (e.g., human users and nodes) to interact with the DLT. In a sense, these software applications are above the DLT core and do not fully belong to the DLT. However, these applications should be developed based on the boundary conditions and rules defined by the blockchain~\cite{rauchs2018distributed}. Risks at this layer affect the functionality and usability of DLT applications. They include risks such as false data feeds, software bugs, and vulnerabilities in user interfaces, which can lead to misinformation, system malfunctions, and unauthorized access, affecting the overall user experience and trust in the DLT application \cite{homoliak2019security}.

\noindent\textbf{External layer:} This layer is at the surface of DLT and includes external additions to the DLT, such as oracles or cryptocurrency exchanges. These risks originate from external factors surrounding the DLT ecosystem. They include regulatory risks, interoperability challenges, market volatility, and adoption barriers, which can impact the legal compliance, integration capabilities, market stability, and acceptance of DLTs within the broader economic and regulatory landscape \cite{homoliak2020security}.

These risks arise from vulnerabilities within the blockchain system at any layer of the stacked model. They can be categorized depending on their origin (for example, network layer risks like denial-of-service attacks or routing attacks), their target (such as consensus layer risks like 51\% attacks or protocol deviations), their impact (like execution layer risks such as privacy breaches or smart contract glitches), or their scope (such as application layer risks like false data feeds or censorship) \cite{HowtoSec14:online}. By categorizing DLT risks based on their sources, impact, target, and scope, stakeholders can develop targeted risk assessment strategies, including network monitoring, data encryption, smart contract auditing, user education, regulatory compliance, and industry collaborations, to mitigate these risks effectively and ensure the security, integrity, and adoption of DLTs \cite{homoliak2020security}.


\section{Research Approach}
\label{RESEARCHMETHOD} 

\subsection{Research Objective}

This study presents a method for assessing the risks of DLT-based applications that organizations can utilize during their development process. The method assists organizations in identifying potential risks and improving the aspects of the DLT-based application to mitigate them. Usually, organizations do not consider risks at the beginning of the development process since they may not be well understood or challenging to assess. In the next section, we will discuss the motivations for conducting this risk assessment.

1) The studies that refer to security in DLTs defined a limited set of attacks on DLTs. Hence, the definition of attacks and potential risks in DLTs are scattered, and most studies have not described a countermeasure for each attack. Therefore, there is a need for a knowledge base documenting all potential attacks for the DLTs and suggesting corresponding countermeasures. In this study, we create a knowledge base for DLTs that contains over 200 attacks. 
Hence, no risk assessment methods in the literature have been developed for DLT-based applications that include a knowledge base of risks that may occur in DLT-based applications. Accordingly, developing a risk assessment method to support the software development life-cycle of such applications is essential. 

2) Distributed ledgers have become increasingly popular over these years due to their suitability to be used in many distributed application scenarios~\cite{serena2022security}. The convergence of security in a DLT still presents many challenges that need to be addressed to take full advantage of the benefits offered by DLTs~\cite{deshpande2017distributed}. Additional research and industry developments are necessary to gain a thorough understanding of the various security aspects of DLT for organizations to effectively address security issues during the DLT development process in a timely manner.
Some key security aspects to consider when working with DLTs include consensus mechanisms, cryptography, identity management, smart contracts, and network security. So, a comprehensive approach to security is critical when working with DLTs, and organizations must carefully consider and address each of these security aspects to ensure the safety and integrity of their data and transactions. Furthermore, the organizations using DLT-based applications are not security experts. So they must continuously acquire volatile knowledge of potential risks and keep themselves updated. Attacks and potential risks in DLTs are scattered; thus, identifying risks specific to DLTs is difficult. Hence, we provide a security risk assessment method for DLT-based applications to help organizations ensure security in the development process.

\subsection{Research Questions}
The main goal of this study is to design a method to assess cyber security risks for DLT-based applications. To achieve our goal, we aim to answer the following research questions:

\noindent\textbf{$RQ_1$:} What are existing security risk assessment methods for DLT-based applications?\\
\noindent\textbf{$RQ_2$:} What cyber security risks are specific for DLT-based applications?\\
\noindent\textbf{$RQ_3$:} What method engineering technique can be used to develop the security risk assessment method?\\
\noindent\textbf{$RQ_4$:} How does the effectiveness of the proposed method?

\subsection{Research Method}

We employed a semi-systematic literature review (SLR) using the snowballing method, combined with method engineering and case study research to design and evaluate the method, collect data to create the knowledge base, and answer the research questions.

\begin{table}[]
\centering
\caption{An overview of the three research methods used in this study with their corresponding research questions (RQ)}

\begin{tabular}{@{}llllll@{}}
\toprule
\textbf{Research Method}     & RQ1 & RQ2 & RQ3 & RQ4  \\ \midrule
Literature study &\checkmark     & \checkmark       &  \checkmark   &          \\
Method Engineering           &        &     &\checkmark     &         \\
Multiple Case Study                   &         &     &     &\checkmark          \\ \bottomrule
\end{tabular}
\label{tab:rm}
\end{table}

\vspace{0.5em}

\noindent\textbf{1) Literature research:} The overall goal of the literature study is to understand the technologies and context of DLT, which is necessary for discovering cybersecurity risks. Furthermore, it establishes a comprehensive overview of what cyber security entails and what methods currently exist. The literature study assists in the method engineering section to create a knowledge base, as it constructs the foundation for the risk database, which is a part of our proposed method. We created an extraction form to collect knowledge, ensure it is consistent with relevant knowledge, and check that the knowledge gathered answered the research questions. The literature study was performed based on the snowballing suggested by Wohlin. Wohlin~\cite{wohlin2014guidelines} presents several guidelines for this method which will be used during the literature study. Because by using snowballing in the literature study process, there is a higher chance of finding research that was published in less-known journals and conferences. Because it reduces the problem of inconsistent terminology, and the critical benefit of snowballing is that it focuses on the cited or referenced papers, which reduces the noise compared to the database approach~\cite{wohlin2014guidelines}. Researchers may use different terminology for the same concepts, and by relying solely on search engines, relevant research may not be found by using deviating terminology. Snowballing circumvents this problem. Hence it can be a suitable method for the DLT context because, in this domain, concepts are often used interchangeably. We aim to answer RQ1 and RQ2 With this literature study.

\noindent\textbf{2) Method engineering:} We use the method engineering to answer RQ3. Based on \cite{brinkkemper1996method}, "Method engineering is the practice of engineering to design, construct and adapt methods, techniques, and tools for developing information systems". It is used to visualize the design process of the method and to engineer it thoroughly. The primary goal of method engineering is to improve the overall efficiency and effectiveness of developing information systems by providing a set of best practices and guidelines for managing the developing process. A Process Deliverable Diagram (PDD) is used to visualize the design and development. It illustrates the activities and artifacts of a specific process. A PDD can be seen as a combination of a business process model and data model~\cite{weerd2009meta}. We will discuss the steps to design our proposed method based on the formal approach for method engineering~\cite{hong1993formal} in section \ref{Analysis}.

\noindent\textbf{3) Case study} research~\cite{jansen2009applied,farshidi2021decision} is an empirical methodology that examines a phenomenon within the context of the particular empirical domain. Interviews are one of the key techniques in case study research to gather data regarding a particular phenomenon or to evaluate a tool with respect to its efficiency and effectiveness. A case study is a widely used and accepted empirical approach for developing theory and conceptual models in the Information Science literature~\cite{pervan2005designing}. This study employs multiple case designs, examining multiple real-world DLT-based applications as multiple cases within their context to evaluate our proposed method. Furthermore, our work has been evaluated using the ACM SIGSOFT Empirical Standards 
\footnote{Find the database through the following link: (\nolinkurl{https://github.com/acmsigsoft/EmpiricalStandards})\vspace{1 em}} for case studies and case synthesis. We used the case studies to answer RQ4.

We conducted a literature study to validate our research and evaluated our proposed method through three case studies. Following the case study methodology of Runeson and Höst~\cite{runeson2012case}, we performed five major steps as follows:

\noindent\textbf{ (1) Case study design}: Define objectives and plan the case study. Building a valid security risk assessment method for the DLT-based application was the primary goal of this research.  \\
\noindent\textbf{ (2) Preparation for data collection}: Procedures and protocols that govern the data collection conduct. The analysis units were three industry case studies performed in the Netherlands and Iran. To conduct the case studies and evaluate the proposed method, we followed the protocol in Appendix\ref{Appendix}\\
\noindent\textbf{ (3) Collecting evidence}: Execution with data collection on the studied case. We conducted multiple expert interviews with the case study participants to understand their requirements, concerns, and preferences regarding the security risk assessment method. \\
\noindent\textbf{ (4) Analysis of collected data} : Data analysis is conducted differently for quantitative and qualitative data. We analyzed the outcomes and observations of the case studies and indicated the most harmed asset for each case study that needs to emphasize safety to mitigate the risk.\\
\noindent\textbf{ (5) Reporting}: The report communicates the findings of the study but is also the primary source of information for judging the quality of the study. We reported the outcomes to the case study participants and received their feedback on the result.


\nopagebreak

\begin{table*}[h!]
\scriptsize
{\renewcommand\arraystretch{1.5}}
\centering
\footnotesize
\caption{shows three risks that have been extracted from the knowledge base \cite{DLTdatbase:online}.}

\begin{tabular}{|m{4em}|m{4em}|m{15em}|m{5em}|m{4em}|m{8em}|m{8em}|m{4em}|m{2em}}
\rowcolor{DarkGray}

\hline
\textbf{Name}         
& \textbf{Synonyms}                  & \multicolumn{1}{l|}{\textbf{Short Description}}    
& \textbf{Harmed asset}  
& \textbf{Impacted Layer} 
& \textbf{Contributes to}  
& \textbf{Relates to}
& \textbf{Reference}
 \\ \hline
 \rowcolor{LightGray}

Double Spending Attack & -  & When the same single digital tokens can be spent on more than one because the attacker manages to spend both of them simultaneously. Double spending is one of the major security issues in most blockchain systems, but it
is difficult to launch unless an adversary has massive computing power successfully.                                                                                            & Network  & Network Layer & - & Finney attack; race attacks; 51 \% attacks   &~\cite{karame2012double}          \\ \hline
 \rowcolor{LightGray}

Eclipse Attack & -
 &  a set of malicious, colluding overlay nodes arranges for a correct node to peer only with coalition members. If successful, the attacker can mediate most or all communication to and from the victim. By supplying biased neighbor information during normal overlay maintenance, a modest number of malicious nodes can eclipse a large number of correct victim nodes.                                                                                                                                                  & Network  & Network Layer  & Consensus Majority Attack, Engineering block races, Splitting mining power (51\% attack), selfish mining, ﻿0-confirmation double spend, ﻿N-confirmation double-spend   & Sybil attack &~\cite{singh2006eclipse}                                  \\ \hline
  \rowcolor{LightGray}

Long Range Attack & History Attack& Proof of Stake vulnerability, due to weak subjectivity and costless simulation, an attacker attempts to rewrite the history of a ledger by forking the ledger
 & Transaction information & Consensus Layer & - & - &~\cite{roy2018inaudible}                 \\ \hline
 
\end{tabular}
\label{tab:riskdatabasetable}
\end{table*}

\section{Systematic literature review}\label{RELATEDWORK}

\begin{figure}[h!]
\centering
\includegraphics[width=0.5\textwidth]{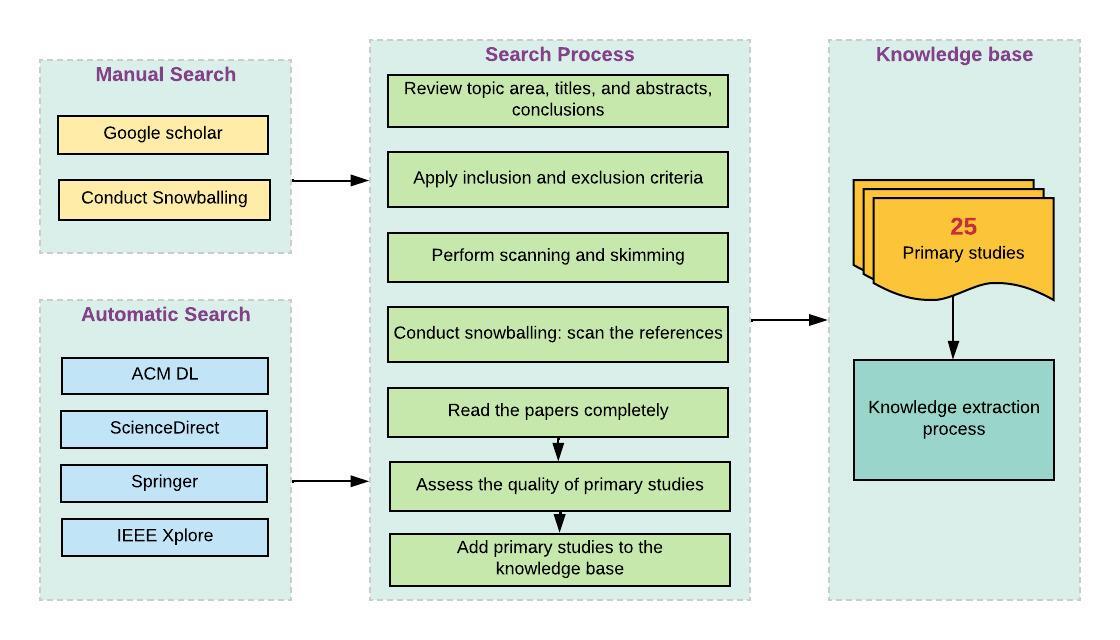}
\caption[SLR]{This figure indicates the stages of the search process.}
\label{fig:SLR}
\end{figure}

As mentioned in~\ref{RESEARCHMETHOD}, we have performed a semi-systematic literature review (SLR) as the primary method to investigate the existing literature regarding security risk assessment methods in different domains. The SLR was performed using the guidelines proposed by~\cite{kitchenham2004procedures}. Moreover, we have also considered a systematic literature review performed by~\cite{FARSHIDI2020110714} as a reference when conducting our SLR. We performed SLR in the following phase:

\noindent\textbf{Data sources and search strategy:}
Generally, it consisted of two search strategies: collection of the initial hypothesis and automatic search. Figure~\ref{fig:SLR} shows the stages of the search process. We gathered a set of papers to generate search terms from their generic keywords during the initial hypothesis search. The search terms are used in the automatic search to collect data from digital libraries. Primary sources for our search are digital libraries, including ACM Digital Library, IEEE Xplore, Springer, and Science Direct.
We have primarily focused on these four libraries since they offer high-quality papers which are valuable in the scientific community.
During the initial hypothesis search phase, we explored literature based on the following search keywords: "security risk assessment method", "risk assessment for DLT", "distributed ledger technology", and "Cybersecurity standards in risk assessment methods", "cybersecurity risk assessment method". Accordingly, We collected a set of papers based on the snowballing method during this phase. Hence, we found 72 papers for risk assessment methods and frameworks with different activities and features.
In order to generate the search terms to use in the automatic search, we extracted the keywords of the papers and employed them in this process. The search terms are: 
\\
\textit{(``risk assessment'' OR ``security risk assessment'' OR ``risk assessment framework'' OR ``assessment framework'' OR ``risk assessment method'' OR ``risk assessment approach'' OR ``risk assessment model'') AND (``blockchain technology'' OR ``distributed ledger'' OR ``distributed ledger technology'' OR ``ledger technology'' OR ``blockchain-based application''). \\}

Afterward, the search term can be used to find publications from digital libraries. Typically, the search results can be exported easily as CSV and BibTex formats. We employed automatic search based on these search terms and automatically extracted 425 manuscripts from the mentioned academic sources. So, in two search methods ( manual and automatic search), we arrived at 495 studies in this phase \\

\noindent\textbf{Inclusion/Exclusion criteria:} Inclusion and exclusion criteria ensure that relevant manuscripts are included and irrelevant manuscripts are excluded. At this phase, based on~\cite{meline2006selecting}, we emphasized excluding studies that clearly meet the exclusion criteria, including titles and abstracts. If the titles and abstracts clearly disqualify them, the studies are eliminated from the candidate studies. 

We extracted the required information, including title, keywords, abstract, the venue where the paper was presented, the number of citations, year, and also the relevancy of each paper with the research questions. 
Thus, based on this information, we have ranked the studies using four qualitative values: None, low, medium, and high. ``none'' indicates that the study is not valuable to our research, and "high" indicates that the study is highly relevant to our research. Finally, we considered 86 studies in this phase.

\noindent\textbf{Quality assessment:}
After the inclusion and exclusion criteria, it is necessary to evaluate the quality of selected papers~\cite{kitchenham2004procedures}
We read and discussed to eliminate concerns and develop consistent criteria for the study quality assessment. Our quality assessment focused on the following questions: 

\begin{enumerate}
    \item Does the study contain a clear problem statement?
    \item Does the study contain research questions? 
    \item Are the clear research challenges of the study mentioned?
    \item Are the results and findings of the research mentioned clearly and understandably?
    \item Does the study contain a real-world use case?
\end{enumerate}

We analyzed and discussed all the studies to eliminate or consider for the next phase. So, we read the full text and eliminated 18 manuscripts based on the above quality assessment criteria. 
The first and second authors conducted a quality assessment of all the results. In this round, 36 studies were eliminated that focused on the risks and attacks in different domains because we considered the studies that designed a security risk assessment method.
Finally, we used 32 studies in our data analysis.\\

\noindent\textbf{Data extraction:}
We identified 25 studies from various domains through a systematic literature review. The results, presented in \ref{tab:slr}, include information on the research method, evaluation technique, risk domain, risk type, and the standards used in each study. Our analysis indicated that two studies presented a risk assessment method exclusively for DLTs. Moreover, we found that the majority of the methods employed in the previous studies were based on ISO 27005 and NIST SP-800-30 standards. To design our proposed risk assessment method, we also considered these standards. Additionally, we were able to answer RQ1 and RQ2 by leveraging the results gained from the SLR.
ISO 27005 is a widely recognized standard for information security risk management. It provides a structured and systematic approach to identifying and assessing risks and developing and implementing risk treatment strategies \cite{al2019risk}. By incorporating ISO 27005 into our proposed method, we can ensure that the risk assessment process is comprehensive, consistent, and aligned with industry best practices.
NIST SP-800-30 is a similar standard that guides conducting risk assessments for information systems. It offers a step-by-step approach to identifying, analyzing, evaluating risks, and developing and implementing risk management plans. By considering this standard in developing our method, we can benefit from the best practices and insights provided by NIST, which is widely recognized as a leading authority in information security.
By combining the insights and best practices from both ISO 27005 and NIST SP-800-30, we can propose that the risk assessment method is likely to be effective, comprehensive, and aligned with industry best practices.

\nopagebreak

\begin{table*}[h!]
{\renewcommand\arraystretch{1.5}}
\centering
\small
\caption{An overview of the results of the SLR}
\begin{tabular}{|p{.6cm}|p{.6cm}|p{3.2cm}|p{2cm}|p{2.5cm}|p{1.3cm}|p{3cm}|}
\rowcolor{DarkGray}

\hline
\textbf{NIST}         
& \textbf{Year}                  & \multicolumn{1}{l|}{\textbf{Research Method}}    
& \textbf{Evaluation Technique}  
& \textbf{Risk Domain}  
& \textbf{RS Type}
& \textbf{Derived}
 \\ \hline
 \rowcolor{LightGray}

\cite{yazar2002qualitative} & 2002 & Literature study  & - & Information security  & Qualitative   & NIST         \\ \hline
\cite{aagedal2002model} & 2002 & Literature study  & Case study & -  & Qualitative   & -     \\ \hline

\cite{asosheh2009new} & 2009 & Literature study & - & -  & Qualitative   & ISO 27001     \\ \hline

\cite{albakri2014security} & 2014 & Literature study & Experiment & Cloud computing & Qualitative   & ISO 27005     \\ \hline

\cite{beckers2014isms} & 2014 & Literature study, 
Benchmarking & Benchmarking & Smart grid & Qualitative   & CORAS,
ISO 27001     \\ \hline

\cite{jagannathan2015cybersecurity} & 2015 & Literature study, 
Benchmarking & Benchmarking & Medical devices & Qualitative   & 
ISO 27001     \\ \hline

\cite{sihwi2016expert} & 2016 & Literature study, 
Benchmarking & Benchmarking & Information system security & Qualitative   & 
ISO 27002     \\ \hline

\cite{cherdantseva2016review} & 2016 & Literature study, 
Benchmarking & Benchmarking & SCADA systems & Qualitative   & 
ISO 31000,
NIST SP 800-30   \\ \hline

\cite{morganti2018risk} & 2018 & Document Analysis, 
Literature study  & Benchmarking & Blockchain & Qualitative   & 
NIST SP-800-30    \\ \hline

\cite{vermeij2018creating} & 2018 & Design science,
Literature study   & Case study & DLT applications & Qualitative   & 
ISO 27005   \\ \hline

\cite{cha2018data} & 2018 & Literature study  & -& Information & Qualitative   & 
ISO 27005,
NIST SP-800-30  \\ \hline

\cite{radanliev2018future} & 2018 & Literature study   & - & IoT & Quantitative   & 
ISO 31000,
NIST SP 800-30 \\ \hline

\cite{weil2020risk} & 2019 & Design science,
Literature study & -& Cloud & Qualitative   & 
ISO 31000,
NIST SP 800-30 \\ \hline

\cite{al2021cyber} & 2021 & Document Analysis,
Literature study & -& Blockchain & Quantitative   & 
ISO 27005,
ISO 31000\\ \hline

\cite{wang2021systematic} & 2021 & Document Analysis,
Literature study & -& Automotive Cybersecurity & Quantitative   & 
ISO 26262\\ \hline

\cite{wang2021cybersecurity} & 2021 & Fuzzy set theory,
Literature study &  Experiment & Control System & Quantitative   & 
ISO 26262\\ \hline

\cite{leszczyna2021review} & 2021 & Literature study,
Benchmarking &  Benchmarking & - & Quantitative   & 
-\\ \hline

\cite{chatzigiannis2021sok} & 2021 & Literature study,
Benchmarking &  Benchmarking & Funding & Quantitative   & 
-\\ \hline

\cite{fucoras} & 2022 & Literature study,
Benchmarking, 
Case study &  Case study & Information security  & Qualitative   & 
ISO/IEC 27005,
CORAS\\ \hline

\cite{bendicho2022cyber} & 2022 & Document Analysis,
Literature study ,
Benchmarking &  Benchmarking & Cloud  & Qualitative   & 
NIST SP 800:30,
ISO 27001,
SO 30001\\ \hline

\cite{angelini2022methodology} & 2022 & Literature study,
Benchmarking, 
Case study &  Case study & -  & Quantitative   & 
SO 27001,
NIST SP 800:30\\ \hline

\cite{durakovskiy2022security} & 2022 & Literature study &  -& Distributed Ledger & Qualitative   & 
ISO 27005,
ISO 27001\\ \hline

\cite{zhang2022sustainability} & 2022 & Literature study,
Benchmarking &  Benchmarking & Blockchain & Qualitative   & 
-\\ \hline

\cite{hossain2022cyber} & 2022 & Document Analysis,
Literature study &  Benchmarking & SCADA system & Quantitative   & 
-\\ \hline

\cite{al2022assessing} & 2022 & Document Analysis,
Literature study ,
Benchmarking &  Benchmarking & IT infrastructure  & Qualitative   & 
-\\ \hline
 
\end{tabular}
\label{tab:slr}
\end{table*}

\section{Security Risk Assessment Method for DLT-based Applications (SRAMDA)}\tabularnewline
\label{SRAMDA}

\begin{figure*}[htbp]
\centering

 \includegraphics[trim=0 0 0 0,clip,width=0.65\textwidth]{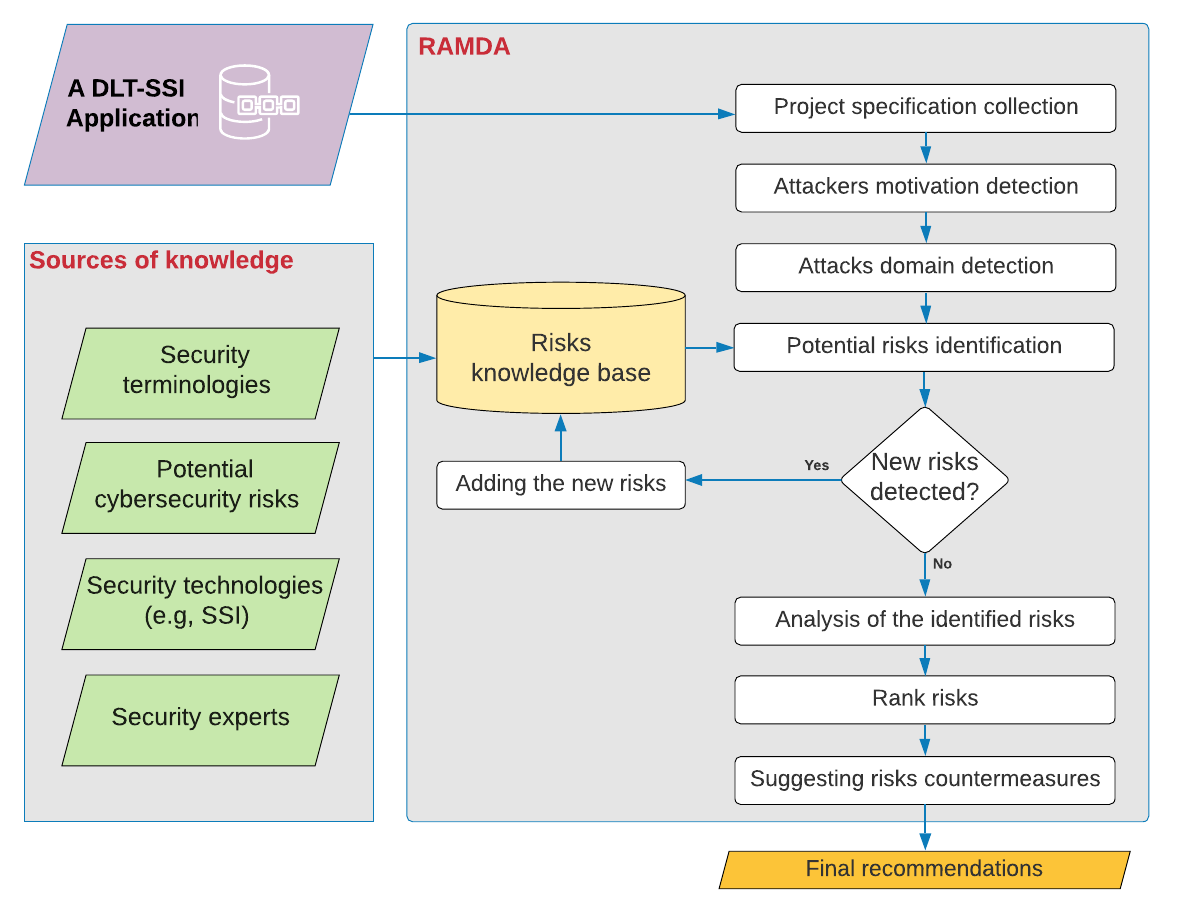}
\centering
\caption[MethodPDD]{The Step-by-step Process of SRAMDA}
\label{fig:MethodPDD}
\end{figure*}

Following the method engineering approach, we propose a simple and holistic approach to risk assessment of DLT-based applications. We call this approach Security Risk Assessment Method for DLT-based Applications, abbreviated as SRAMDA. Figure~\ref{fig:MethodPDD} illustrates the SRAMDA process. The method relies on the existing knowledge that has to be collected before the core risk assessment steps. Therefore, risk analysts following SRAMDA need to 
\begin{enumerate*}
   \item analyze security terminology,
   \item conduct DLT analysis, and
   \item gather data-related risk.
\end{enumerate*}
In the latest step, analysts could use our DLT knowledge base which currently contains 115 risks specific to DLT-based applications. The description of the risks in SRAMDA knowledge base provides information about the name of the attack, description, related risks, possible harmed assets, impacted layer (based on stack model of DLT), and references to the source of information about the attack~\cite{DLTdatbase:online}. However, we indicated whether each attack was specifically targeting DLTs or if it belonged to well-known cybersecurity attacks. Based on our findings, 28.95\% of the attacks were identified as common cybersecurity threats, while a significant majority of 71.05\% were explicitly tailored to exploit vulnerabilities within DLTs. It was imperative for us to maintain the integrity of our examination by providing valid references for each of the included attacks, thereby ensuring the validity and reliability of our research. Table~\ref{tab:riskdatabasetable} demonstrates a fragment of the knowledge base.



Once the knowledge base preparation is done, the SRAMDA guides analysts into the core risk assessment steps. This part consists of ten steps which we describe in what follows:

\paragraph{Project specification collection}    
The first step is gathering project specifications, or 'context establishment' as known in ISO 27005. This step requires risk analysts to understand the project's specifics and discover the `crown jewels' to be protected. The risk analyst and stakeholders must define common terminology and decide the scope boundaries in order to focus on the future risk assessment and make it feasible. This step aims to set common ground and reduce possible misunderstandings between stakeholders and risk analysts. 

\paragraph{Attackers motivation detection}
Based on the project description and scope produced in the previous step, the risk analyst can analyze possible cyber criminals' motivations for attacking the system.  This risk assessment assumes an opportunity to attack the system under analysis, and the motive is the specific type of attack. With early design and development systems, performing a risk assessment method can be difficult because specific components are undefined. However, one can reason why an attacker would want to attack this specific system. For example, if there are monetary assets. This eliminates some attacks, as their probability will be very low. The main objective of this step is to determine the motive behind an attack on the system under analysis, such that the risk analyst can decide what risks are most likely to be exploited. 

We adopted the following classification of the attacker's motivations: 
\begin{itemize}
   \item[] \textbf{Monetary:} The probably most apparent extrinsic motivation for an attacker is the financial benefit the attack can result in. An attacker could, for instance, commit fraud or extort other people with the help of cyberattacks. 
   \item[] \textbf{Damage:} Causing damage is another extrinsic motivation that an attacker may have. The willingness to cause damage can come from various reasons.
   \item[] \textbf{Knowledge:} The other motivation is an attacker's knowledge from performing an attack and how advantage is reached through the theft of information and essential data.
   \item[] \textbf{Trust:} An attacker interested in notoriety within a community intends to destroy trust.
\end{itemize}

\paragraph{Attacks domain detection}
In this step, the risk analyst determines the domain of interest for attackers by annotating each motivation. The risk analyst should determine where the attacker would seek motivation to get his desired asset. 
For example, if the attack's motive is monetary assets, they might attack the layer in which the value exchange occurs. In this step, the risk analyst identifies the risks based on the knowledge base and which assets might be vulnerable. 

\paragraph{Potential risks identification}
In this step, the risk analyst can identify potential risks by comparing the information earned from previous steps with the knowledge base. In previous steps, the motivations of the attacker and the domain of interest (type of motivations) were determined. We collected 226 attacks in the DLT-based applications and determined the DLT layer for each attack and the type of attacks (Monetary, Damage, Knowledge, and Trust) in the knowledge base. Hence, the risk analyst can filter the attacks in the knowledge base based on these data.

   \subsubsection {New risks detected}
   This step is unique compared to most risk assessment methods. As mentioned before, this method utilizes a risk knowledge base that was earned during an extensive literature study. The knowledge base consists of all relevant identified risks on DLT.

If the risk analyst reviews the knowledge base and does not identify any risks or attacks based on the information gathered from previous steps. In that case, they can explore the possibility of discovering new attacks by conducting a literature study focused on identifying potential risks in DLT-based applications.
\\

   OR-gateway:
\begin{itemize}
\item[]     If \textit{yes}: Add new risks to the knowledge base, then continue to the next step.
\item[]     If \textit{no}: Continue to the next step. 
\end{itemize}

\paragraph{Analysis of the identified risks}

The risk analyst must identify and describe threats based on the identified assets. In this step, the risk analyzer describes the attack scenarios (Events produced by the attacker to attain its attack goal) leading to the achievement of the attack goals determined in the second step to estimate the probability of occurrence of threats.

\nopagebreak

\paragraph{Rank risks}
  
   The identified and analyzed risks are reported to the project leaders in this step. If they confirm that the risks can be potential threats and attacks on their project, the risk analyst asks them to rank the risks because the risk analyst might not be capable of assessing the severity of a risk. This step's objective is to rank the identified and evaluated risks and determine what risks are essential in the project to take into account. Also,  it eliminates irrelevant risks and diminishes the number of treatments that have to be presented for the next step.
    
\paragraph{Suggesting risks countermeasure}
    
     The result of this step should be a structured list of possible risk treatment and countermeasure options per risk. Countermeasures can be collected from scientific literature, querying search engines, asking security experts, or using their own experiences and knowledge of the analyst. Risk treatments are not included in the database by default, as not every treatment might work on each project. Treatments are very project-specific; hence it is dangerous to blindly assume that a specific treatment in a database will completely mitigate risk without acquiring more understanding. Consequently, it might be best to let the analyst survey the treatment description intensively to conclude what treatment options might be most relevant and effective.  

\paragraph{Final recommendation}
    This step is the last step that has to be documented. Here, the risk analyst concludes the risk assessment and provides the project leaders with her final recommendation. The recommendations are presented to the project leaders based on the most vulnerable asset. The risk analyst checks the number of identified risks that are detected based on the harmed asset in Step 6. Then the risk analyst provides the mitigation and prevention operation against these risks as final recommendations.


\section{Empirical Evidence: The Case Studies}\label{CaseStudies}

We aimed to evaluate the effectiveness of SRAMDA through three case studies. To ensure a diverse evaluation, we selected case study organizations from three different domains: data security (SecureSECO), financial services (Mobifi), and crypto exchange (Aratoo). These organizations were located in the Netherlands and Iran. We specifically chose companies as our case studies that were in the process of developing DLT-based applications.

\subsection{SecureSECO}

In this section, we outline all the steps of SRAMDA that were conducted for the SecureSECO project.Figure~\ref{fig:SecureSECO} shows the summary of employing these steps.

\subsubsection{Gather project specifications}

SecureSECO aims to secure and increase trust in the software ecosystem through DLT and empirical software engineering research. SecureSECO aim to create a secure software ecosystem. This project uses DLT to store valuable trust facts about software and users. This project relies on trust, which is hard to measure, but it is possible to protect and use smart contracts and consensus mechanisms. The project leaders determined the scope of risk assessment during the interview as follows: \textit{To provide valuable insights and awareness of cybersecurity risks. It can provide the assessment results for further designing and developing the project.}

\begin{table*}[htbp]
\centering
\caption{ A step-by-step application of SRAMDA to SecureSECO.}

 \includegraphics[trim=13 210 13 50,clip,width=1.0\textwidth]{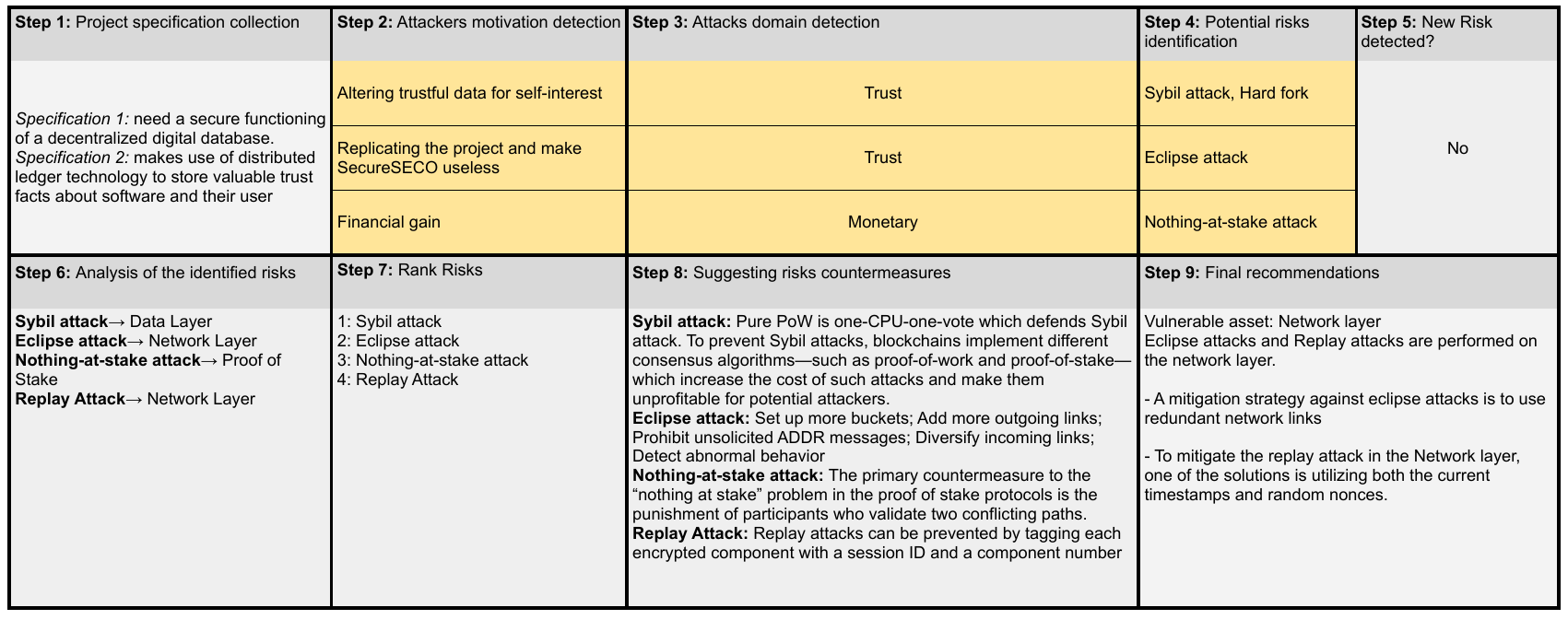}
\centering
\label{fig:SecureSECO}
\end{table*}

\subsubsection{Attackers Motivation}
The motivations of an attacker are as follows:

\noindent\textbf {Altering trustful data for self-interest:} For this motive, the attacker seeks to disrupt the system completely. This will probably be done with a direct attack. Altering the trust facts can lead to discrediting SecureSECO.
\\
\noindent\textbf{Replicating the project and making SecureSECO useless:} An attack could have the motive of replicating the project and making the original SecureSECO redundant. For example, The attacker could monetize the new project and add more data than in the original project. Furthermore, A group of attackers could have the motive of taking over the network, either shutting it down, poisoning the well or having other malicious activities on the network.
\\
\noindent\textbf{Financial gain:} obtaining salable information, either resell personal details or the trust facts.

\subsubsection{Attacks domain detection}

In this step, we determined the domain of interest for attackers by annotating each motivation (see Table~\ref{fig:SecureSECO}).

\subsubsection{Potential risks identification}

This step identified potential risks by comparing the information earned from previous steps with the knowledge base. We identified the following attacks for SecureSECO:

\noindent\textbf{Potential Risk: Sybil attack:} A Sybil attack is one where an attacker pretends to be so many people simultaneously. (Attackers create multiple virtual identities)~\cite{levine2006survey}.

\noindent\textbf{Potential Risk: Eclipse attack:} Eclipse attacks are network attack that aims at eclipsing specific nodes from the entire P2P network~\cite{heilman2015eclipse}.

\noindent\textbf{Potential Risk: Nothing-at-stake attack:} Validators motivate to operate on various forks, notwithstanding the protocol diversity of PoS. Validators could make conflicting blocks on the probable forks with nothing at stake. This problem is called the nothing at stake attack, which reduces the consensus time of the network, lessens the system efficiency, and decreases the blockchain capability to solve double-spending attacks and other threats~\cite{rebello2021security}.

\noindent\textbf{Potential Risk: Hard fork:} When a system splits into two systems and a new version is created, which is not compatible with the old version, as the old nodes could not agree with the mining of the new nodes: consensus cannot be reached. A hard fork is not reversible. ~\cite{heilman2015eclipse}.

\subsubsection{New risks detected?}
We did not identify new risks for SecureSECO and found all attacks based on our knowledge base. So we can continue to the next step.

\subsubsection{Analysis of the identified risks}

In this step, we described the attack scenarios for SecureSECO based on the information earned from previous steps (See Table~\ref{fig:SecureSECO}).

\subsubsection{Rank risks}
In this step, we reported the identified and analyzed risks to the SecureSECO project leaders. They confirmed that the risks could be potential threats and attacks for SecureSECO, and they ranked the attacks for their project to indicate the prioritization and importance of risks. 

\subsubsection{Suggesting risks countermeasures}
In this step, we prepared a list of possible risk treatment and countermeasure options per risk.

\noindent\textbf{Countermeasure: Sybil attack:} Pure PoW is one-CPU-one-vote which defends Sybil attack~\cite{wright2008bitcoin}.

\noindent\textbf{Countermeasure: Eclipse attack:} Set up more buckets; Add more outgoing links; Prohibit unsolicited ADDR messages; Diversify incoming links; Detect abnormal behavior~\cite{heilman2015eclipse}.

\noindent\textbf{Countermeasure: Nothing-at-stake attack:} The primary countermeasure to the "nothing-at-stake" problem in the proof of stake protocols is the punishment of participants who validate two conflicting paths~\cite{rebello2021security}. 

\subsubsection{Final recommendations}

Most of the risk occurs in the Network layer in the DLT of SecureECO, which is more vulnerable to attacks. Thus the recommendation to mitigate these attacks can be helpful for their project. Eclipse attacks and Replay attacks are performed on the network layer. A mitigation strategy against eclipse attacks is to use redundant network links or out-of-band connections to verify transactions (e.g., by a blockchain explorer). To mitigate the replay attack in the Network layer, one of the solutions is utilizing both the current timestamps and random nonces. To achieve this goal, all the entities in the network are assumed to be synchronized with their clocks. This is a typical assumption applied in designing various security protocols in networks.

\subsection{Case Study 2: Mobifi}
\subsubsection{Gather Project Specifications}
MobiFi is a financial organizer for Mobility as a Service (MaaS) — which aims to solve this. It is based on a DLT-based payment engine that automates the mobility credit system for enterprises, all managed in one control portal.
They provide a transparent platform with a tokenized payment system to connect mobility service providers and users. MobiFi bridges the gap between the mobility industry and Decentralized Finance (DeFi).

The scope of this risk assessment was determined along with the project leaders during the group interview and is formulated as follows:
\textit{The scope of this risk assessment is to provide valuable insights and awareness of the cyber security of Mobifi, such that it can take the results of SRAMDA into account while further designing and developing the project. }

\subsubsection{Attackers Motivation}

The motivations of an attacker are as follows:

\noindent\textbf{Motivation: Gaining knowledge -} There is significant competition between companies that produce transportation information. For example, telecommunication companies generate data that can be used for transportation modeling. Similarly, the logs of available mobile devices registered by cellphone towers can be used to monitor traffic.

\noindent\textbf{Motivation: Access sensitive data on the network nodes -} Mobility data is continuously generated by different network nodes. According to the identification layer of the multi-layered blockchain model for the Mobifi data market, transportation data is stored in Identification files. Each entity contains metadata, static data, and dynamic data. Nodes need consent from the owner to access the static and dynamic data, which can be attractive to many actors.

\noindent\textbf{Motivation: Sabotage activities -} Cybercriminals may use the blockchain ecosystem for sabotage activities. Since the process is anonymous, it is hard to track user behaviors and be subject to legal sanctions. For example, criminals can leverage smart contracts for various malicious activities, which may threaten our daily lives. Manipulate and modify blockchain information: Blockchains are made to be practically immutable, where, in theory, no one can modify the blockchain's distributed ledger of all committed blocks. The blockchain relies on the distributed consensus mechanism to achieve immutability and establish mutual trust.

\subsubsection{Attacks domain detection}
In this step, we determined the domain of interest for attackers by annotating each of the motivations (See Table~\ref{fig:Mobifi}).

\subsubsection{Potential risks identification}

This step identified potential risks by comparing the information earned from previous steps with the knowledge base. We identified the following attacks for Mobifi:

\noindent\textbf{Potential Risk: Wormhole attack within a channel -} The permissioned blockchain technology is prone to wormhole attack because the sender and receiver identities are not hidden on the channel of a permissioned blockchain as per lack of peer privacy. Within a channel, compromising a member causes to leakage of all members' ledger information to everyone outside the channel. Within a private network, a malicious node creates a virtual private network with the outside network and leaks the information of its private network. This attack can be launched without any knowledge of honest nodes of the private network~\cite{al2021cyber}.

\noindent\textbf{Potential Risk: Node spoofing attack -} Node spoofing is when an attacker steals Decentralized Identifier (DID) credentials exploiting a vulnerability in an improper key protection mechanism and communicates with another node on behalf of the user. If cryptographic keys are not stored or maintained properly, it could cause the compromise and disclosure of private keys leading to fraudulent transactions or loss of assets. This will lead to compromising the integrity and privacy of the operations. Wallet theft uses classic mechanisms such as phishing, including system hacking, the installation of buggy software, and the incorrect use of wallets~\cite{iqbal2019blockchain}. 

\noindent\textbf{Potential Risk: Dictionary Attack -} Attackers use dictionaries of known passwords, a subset of brute force attacks. This can be used against services requiring login or against cryptographically protected data requiring a password or passphrase to access it, such as a wallet~\cite{abdullah2017blockchain}.

\noindent\textbf{Potential Risk: Sybil attack -} Attackers create multiple virtual identities, which means a single faulty entity can present multiple identities, and it can control a substantial fraction of the system, thereby undermining this redundancy~\cite{douceur2002sybil}.

\noindent\textbf{Potential Risk: Credential Stuffing -} Attackers use spilled or otherwise leaked credentials and account names to try name/password combinations with a higher likelihood of success against services requiring authentication~\cite{thomas2019protecting}.

\noindent\textbf{Potential Risk: Time hijacking attack -} Time hijacking attacks occur because of the vulnerability of Bitcoin time stamp processing. The time counter of the Bitcoin network is modified, and nodes' time changes as well. Hardware-oriented system time will be adopted to replace the previous network time. Time hijacking is prone to occur when an adversary executes a Sybil attack with inaccurate timestamps simultaneously, which adds multiple Sybil nodes to the network~\cite{decker2014bitcoin}.
    
\noindent\textbf{Potential Risk: BGP routing attack -} Border Gateway Protocol belongs to external routing protocols connecting different networks on the Internet. In BGP routing attacks, known as BGP hijacks, any malicious AS (Autonomous System) can create fake advertisements for any prefix and advertise them to its neighbors, redirecting traffic directed to given destinations~\cite{wen2021attacks}.

\noindent\textbf{Potential Risk: Malleability attack -} In the Hyperledger network, the ledger of a channel inside the Hyperledger limits the accessibility to only members that are part of the channel. The client can choose an endorser of its choice during the transaction proposal phase, due to which the identity of the endorser is disclosed to everyone in the network, including an insider adversary. Suppose an attacker is a channel member on Hyperledger, as in conventional data-sharing schemes. In that case, the attacker can eavesdrop on all the network traffic inside a channel of Hyperledger fabric by exploiting vulnerability related to lack of peer privacy. The attacker also has access to every transaction present in the ledger~\cite{decker2014bitcoin}.

\begin{table*}[htbp]
\centering
\caption[]{A step-by-step application of SRAMDA to MobiFi.}

 \includegraphics[trim=15 165 15 50,clip,width=1.0 \textwidth]{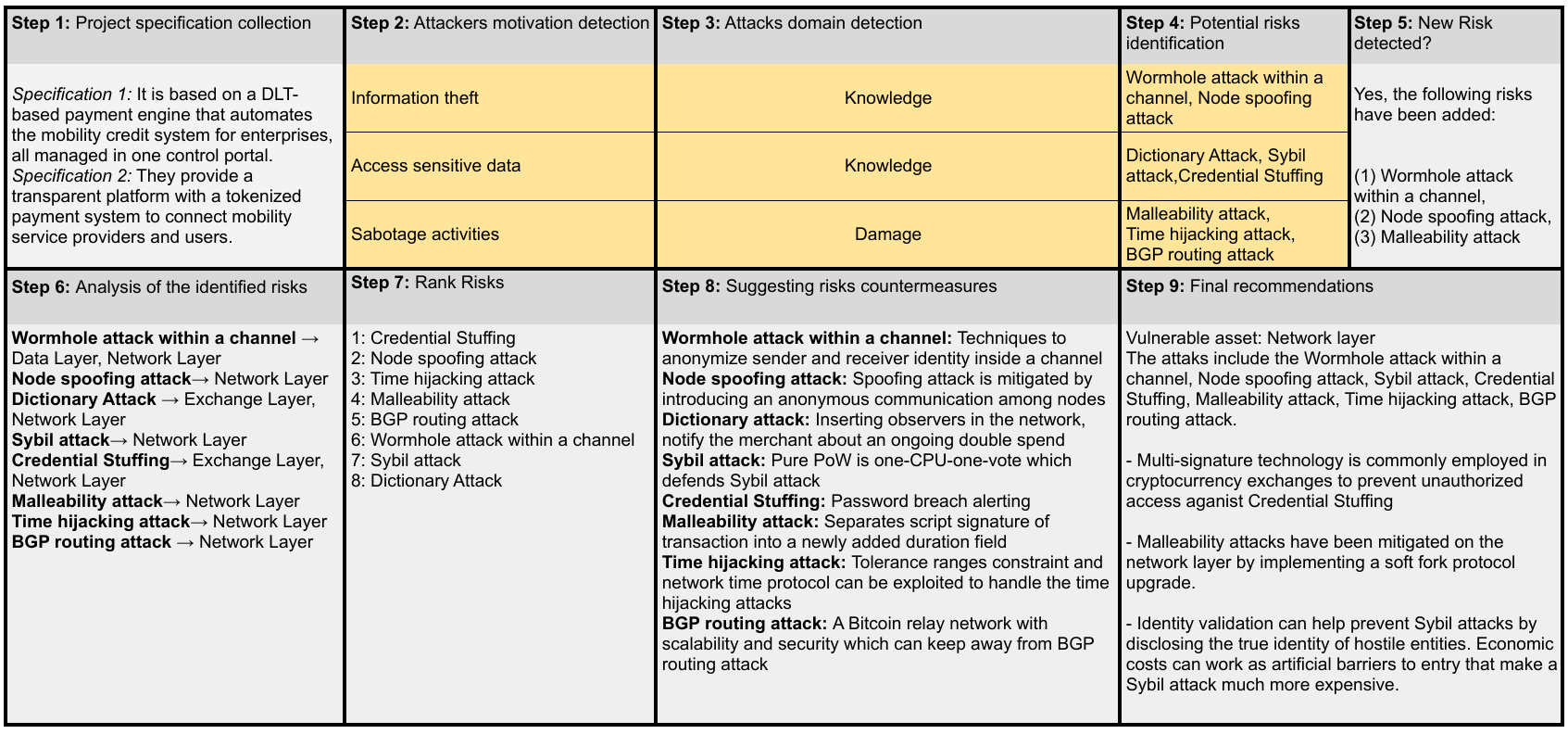}
\centering
\label{fig:Mobifi}
\end{table*}

\subsubsection{New risks detected?}

We needed to perform a literature study for some motivation, So we found new risks. Wormhole attacks within a channel, Node spoofing attacks, and Malleability attacks were new risks we detected based on the literature study. Hence, we added them to our knowledge base.

\subsubsection{Analysis of the identified risks}

In this step, we described the attack scenarios for Mobifi based on the information earned from previous steps (See Table~\ref{fig:Mobifi}).

\subsubsection{Rank risks}
In this step, we reported the identified and analyzed risks to the Mobifi project leaders. They confirmed that the risks could be potential threats and attacks for Mobifi, and they ranked the attacks for their project to indicate the prioritization and importance of risks. 

\subsubsection{Suggesting risks countermeasures}

In this step, we prepared a list of possible risk treatment and countermeasure options per risk.

\noindent\textbf{Countermeasure: Wormhole attack within a channel -} To address this weakness in the consensus design, techniques to anonymize sender and receiver identity inside a channel should be implemented~\cite{al2021cyber}. 

\noindent\textbf{Countermeasure: Node spoofing attack -} Spoofing attack is mitigated by introducing an anonymous communication among nodes~\cite{iqbal2019blockchain}.

\noindent\textbf{Countermeasure: Dictionary attack -} inserting observers in the network, notify the merchant about an ongoing double-spend~\cite{rathod2018security}.

\noindent\textbf{Countermeasure: Sybil attack -} Pure PoW is one-CPU-one-vote which defends Sybil attack~\cite{wen2021attacks}.

\noindent\textbf{Countermeasure: Credential Stuffing -} password breach alerting~\cite{HowdoesB4:online}.

\noindent\textbf{Countermeasure: Malleability attack -} Separates script signature of the transaction into a newly added duration field.

\noindent\textbf{Countermeasure: Time hijacking attack -} Tolerance ranges constraint and network time protocol can be exploited to handle the time hijacking attacks~\cite{wen2021attacks}.

\noindent\textbf{Countermeasure: BGP routing attack -} A Bitcoin relay network with scalability and security which can keep away from BGP routing attack~\cite{wen2021attacks}.

\subsubsection{Final recommendations} After the analysis of Mobifi, we realized that most attacks damage the network layer of their DLT, including Wormhole attacks within a channel, Node spoofing attacks, Sybil attacks, Credential Stuffing, Malleability attacks, Time hijacking attacks, BGP routing attack. Hence, mitigating the mentioned risks in the network layer is essential. 
Credential stuffing is easy to execute on vulnerable systems, which makes it all the more important to protect the organizations and the data of millions of users who trust them. Multi-signature technology is commonly employed in cryptocurrency exchanges to prevent unauthorized access to users' accounts because it requires more than one signature to authorize a process. Malleability attacks have been mitigated on the network layer by implementing a soft fork protocol upgrade. There are several problems a Sybil attack may cause, such as Block users from the network. Identity validation can help prevent Sybil attacks by disclosing the true identity of hostile entities. Economic costs can work as artificial barriers to entry that make a Sybil attack much more expensive. Personhood Validation is another solution to mitigate the Sybil attack. A validation authority can use a mechanism that does not require knowing the real identity of participants.

\subsection{Case Study 3: Aratoo}
\subsubsection{Gather Project Specifications}

Aratoo is an Iranian decentralized autonomous organization that aims to liberate the Iranian cryptocurrency market. They are developing, amongst other products, a DeFi wallet for managing cryptocurrencies. A DeFi wallet is a non-custodial wallet where the users have complete access and control of their private keys and funds. DeFi wallets are at the core of the concept ``be your own bank''. Aratoo is a transparent platform that uses smart contracts and native protocol to reduce investment risk, increase profits, and expand blockchain technology and decentralized systems. They have employed DeFi ecosystems, such as MakerDAO and CurveDAO, for lending, borrowing, exchanging, and governing. The DAO will allow liquidity providers to decide on adding new pools, changing pool parameters, token incentives, and many other protocol aspects. A pool is a smart contract that implements the StableSwap invariant, thereby exchanging two or more tokens. The scope of this risk assessment was determined along with the project leaders during the group interview and is formulated as follows:
\textit{to provide valuable insights and awareness in the cyber security of Aratoo, such that it can take the results of SRAMDA into account while further designing and developing the project.}

\subsubsection{Attackers motivation detection}

The motivations of an attacker are as follows:

\noindent\textbf{Motivation: Financial profit -} These are the operations carried out by the attacker to acquire assets or goals and are not limited to brute-forcing credentials, reconnaissance surveillance, exploit delivery, etc.

\noindent\textbf{Motivation: Access sensitive data -} attacker aims to take over a user account through various attacks and gain access to protected data.

\noindent\textbf{Motivation: Information theft -} This attack aims to obtain a token without the attacker paying for it with its currency. 

\noindent\textbf{Motivation: Acquisition of cryptocurrency:} The attacker needs to obtain somehow the target user's private key or the login ID and password for the user's web or another wallet.

\noindent\textbf{Motivation: Steal information -} The attacker aims to shut down communication between parties or steal information from an unsuspecting victim.

\noindent\textbf{Motivation: Sabotage activities -} The attack targets those ordinary users and exchanges that identify transactions based solely on hashes without checking the contents of the transactions.

\subsubsection{Attacks domain detection}

In this step, we determined the domain of interest for attackers by annotating each motivation (See Table~\ref{fig:SecureSECO}).

\subsubsection{Potential risks identification}

This step identified potential risks by comparing the information earned from previous steps with the knowledge base. We identified the following attacks on Aratoo:

\noindent\textbf{Potential Risk: Cryptomining -} Cryptomining (also known as Cryptojacking) involves an attacker using victims' compute resources to mine cryptocurrencies; this can range from using malware to stolen credentials to gain access to systems~\cite{tuttle2018cryptojacking}.

\noindent\textbf{Potential Risk: Account Hijacking -} Account Hijacking is an attack where a malicious user (attacker) can retrieve a cloud user credential and use it in the attacker's favor~\cite{stodt2021security}.

\noindent\textbf{Potential Risk: Double spending attack -} When the same single digital token can be spent on more than one because the attacker simultaneously manages to spend both of them. Double spending is one of the significant security issues in most blockchain systems, but it is challenging to launch unless an adversary has massive computing power successfully~\cite{zachariadis2019governance}.

\noindent\textbf{Potential Risk: Wallet theft -} Where credentials, such as keys associated with peers in the system, are stored in a digital wallet, the "wallet theft" attack arises with specific implications for the application~\cite{collomb2016blockchain}.

\noindent\textbf{Potential Risk: Replay Attack -} In Blockchain systems, the replay attack is the most common issue faced by Blockchain transactions and can cause a long delay in the communication between the two parties~\cite{anita2019blockchain}. 

\noindent\textbf{Potential Risk: Selfish mining Attack -} Selfish mining is a strategy in which a group of miners cooperate and tactically withhold transactions to enter the ledger and then release them to increase their profits~\cite{natarajan2017distributed}.

\noindent\textbf{Potential Risk: Transaction Malleability -} Transaction malleability is one of the critical threats to the blockchain, which can facilitate double-spending attacks by tampering with the state of a blockchain~\cite{zachariadis2019governance}.

\begin{table*}[htbp]
\centering
\caption[]{A step-by-step application of SRAMDA to Aratoo.}
 \includegraphics[trim=17 160 17 50,clip,width=1.0\textwidth]{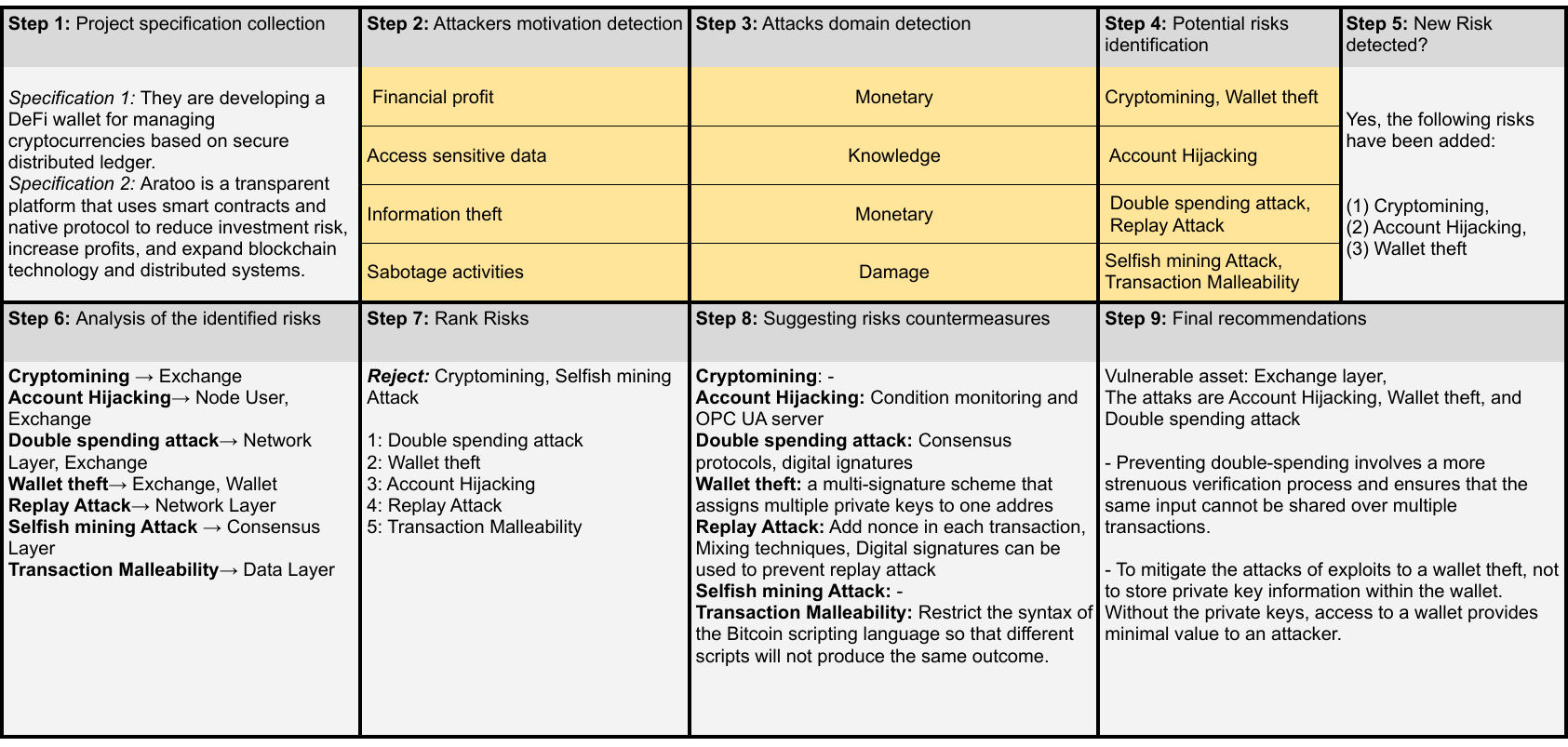}
\centering
\label{fig:Aratoo}
\end{table*}

\subsubsection{New risks detected?}

We needed to perform a literature study for some motivation, So we found new risks. Cryptomining, Account Hijacking, and Wallet theft were new risks that we detected based on the literature study. Hence, we added them to our knowledge base.

\subsubsection{Analysis of the identified risks}
In this step, we described the attack scenarios for Aratoo based on the information earned from previous steps (See Table~\ref{fig:SecureSECO}).

\subsubsection{Rank risks}
In this step, we reported the identified and analyzed risks to the Aratoo project leaders. 
They rejected two risks, including crypto mining and selfish mining attacks. They mentioned that their project's probability of these attacks happening is low. However, they confirmed the rest of the risks and ranked the attacks for their project to indicate the prioritization and importance of risks.

\subsubsection{Suggesting risks countermeasures}

In this step, we prepared a list of possible risk treatment and countermeasure options per risk.

\noindent\textbf{Countermeasure: Account Hijacking -} Condition monitoring and OPC UA server: an attacker uses compromised/stolen credentials to access and impersonate the account. Typically, account hijacking is done through phishing, sending fake emails to the user, password guessing, or various other tactics. By accessing these systems, it is possible to send valid fake data easily. An attacker could also compromise the active session and gain access to the transmitted data~\cite{stodt2021security}.

\noindent\textbf{Countermeasure: Double spending attack -} Consensus protocols, digital signatures.

\noindent\textbf{Countermeasure: Wallet theft -} The Bitcoin system offers a multi-signature scheme that assigns multiple private keys to one address. Under this scheme, the theft of one of the necessary private keys alone would not enable the coins to be stolen unless the rest of the necessary keys are stolen as well~\cite{lindman2017opportunities}.

\noindent\textbf{Countermeasure: Replay Attack -} Add nonce in each transaction; mixing techniques and digital signatures can be used to prevent replay attack~\cite{conti2018survey}. 

\noindent\textbf{Countermeasure: Transaction Malleability -} Restrict the syntax of the Bitcoin scripting language so that different scripts will not produce the same outcome~\cite{stodt2021security}.

\subsubsection{Final recommendation: } As shown in Table~\ref{fig:Aratoo}, most of the risks sabotage the exchange component in the DLT of the Aratoo project. These attacks are Account Hijacking, Wallet theft, and Double spending attacks. Thus the exchange component is vulnerable, and they need to consider the prevention operation against the mentioned attacks. Preventing double-spending involves a more strenuous verification process and ensures that the same input cannot be shared over multiple transactions. To mitigate the attacks of exploits to a wallet theft, not to store private key information within the wallet. Without private keys, access to a wallet provides minimal value to an attacker. 

\section{Analysis}\label{Analysis}

Four research questions were created to give guidance in developing the method. These research questions will be answered and reflected upon in this section.

\noindent\textbf{RQ1: What are existing security risk assessment methods?}

\noindent There are several security risk assessment standards and frameworks for conducting risk assessments in organizations, such as NIST 800-30 and ISO 27005. These standards provide the general and main steps to conduct security risk assessments. NIST SP 800-30 includes a series of practical guidance on information security technology and management issues (NIST, 2002). ISO 27005 (ISO, 2011) is a standard for information security risk management~\cite{cherdantseva2016review}. Many methods have adopted the ISO 27005 standards, which standardized risk assessment methods for industrial use. An example of such a method is the CORAS method~\cite{beckers2014isms} which is a model-based risk assessment methodology for security-critical systems. The ISO27000 method consists of seven steps, from defining the risk assessment context to reviewing, monitoring, and auditing the performed method. It can be concluded that it is hard to perform risk assessments, as they are often probabilistic, and likelihood and consequences are therefore hard to estimate precisely~\cite{cherdantseva2016review}. Most risk assessment methods focus on threat modeling, mitigation, and response planning and are performed for well-defined concepts~\cite{pfleeger2010measuring}. 
We examined other risk assessment methods that provided insight into how risk assessments are performed, and the results helped us as guidelines for the method engineering approach. We reviewed the most popular security risk assessment methods (based on the table~\ref{tab:slr}), such as ISO 27005, and NIST SP800-30, to determine the critical steps in security risk assessment. In this paper, we relied on the ISO27005  and NIST SP800-30 standards to define the main steps of the security risk assessment. 

\noindent\textbf{RQ 2: What cyber security risks are specific for DLT-based applications?}

\noindent There are hundreds of cyber security risks for DLT-based applications. Based on the literature study, we reviewed the studies that introduced attacks and risks for DLTs and blockchain. The most studies employed benchmarking to analyze and compare a collection of risks against each other in literature. Finally, a database could be constructed with 115 risks for DLT-based applications. The process starts with the knowledge-based creation process, which consists of three steps: (i) analysis of security terminology, (ii) analysis of DLT, (iii) the gathering of data-related risks, including the name of the attack, description of the attack, related risks, harmed asset, impacted layer, and a citation~\cite{DLTdatbase:online}. The "Relates to" column in the data set represents the relationships that we found in the literature between different attacks. However, please note that this list is likely to be incomplete, as we only considered a limited set of peer-reviewed and grey literature. Table~\ref{tab:riskdatabasetable} shows a piece of the knowledge base. The DLT risk knowledge base already contains 115 risks for DLT-based applications in the different security areas, and each cyber security risk was related to at least one layer of DLT.

\noindent\textbf{RQ3: What method engineering technique can be used to develop the cyber risk assessment method?}
 
\noindent Based on the formal approach for method engineering proposed~\cite{hong1993formal}, we use the following steps to design our proposed method:\\
1. Method selection: Based on the SLR that we mentioned in~\ref{tab:slr}, we realized that ISO 27005 and NIST SP 800-30 are the most popular standard in the studies. So, we employed their guidelines for risk assessment to determine the critical steps in security risk assessment. \\
2. Method modeling: In order to design and develop a risk assessment method, a process-deliverable diagram is used for creating the meta-model of the method. It is a valuable method for determining the activities and the outputs of the method~\cite{weerd2009meta}. This meta-modeling technique is straightforward and consistent with UML standards (see figure~\ref{fig:MethodPDD}.\\
3. Method development: The method, modeled in PDD, is decomposed into activities and concepts. 
In~\ref{SRAMDA}, we described all activities and concepts of all steps.

\noindent \textbf{RQ4: How does the effectiveness of the proposed method?}

\noindent In order to evaluate the method, we used the case study research and guidelines of Runeson et al.~\cite{runeson2012case}, Yin~\cite{yin2009case} and Robson~\cite{robson2002real}. We selected ten companies that used DLT-based applications in their project and reviewed the documents we found on their website. Our criteria for selecting the companies as a case study are: 1) currently, they have used DLT, or they want to develop their own DLT. 2) We also considered the size of the companies on our list. We checked the size of the companies on our list. The size was different in the range of 10 to 500. We excluded the companies that they had less than 50 employees. Because we wanted to keep the well-known and larger companies on our list. 3) We contacted the companies and asked them if they were interested in participating as a case study in our research. Finally, we ended up with three case studies on our list. Three industry case studies have been conducted to evaluate and signify the risk assessment's efficiency and effectiveness. We conducted a set of interviews with the experts at three case study organizations and asked them a set of questions based on the case study protocol. Then, we analyzed the results of the interviews. Table~\ref{fig:Comparison} shows a comparison between three case studies. \\

\section{Discussion}\label{Discussion}

In this study,  we developed a method to assess the risks for DLT-based applications. A case study was the best suitable method for evaluating the SRAMDA, as the method is designed to be applied to a real-world case. We conducted three case studies and asked them a set of questions based on the case study protocol (See Appendix~\ref{Appendix}) to analyze their project and assess potential risks that they might face during the development.
 For two steps of SRAMDA appeared to be very difficult to determine which of the 115 risks were most relevant to a project still in the proof-of-concept stage. These two steps include the attacker's motivation detection and attack domain detection.
Determining the motivation and attacks domain made it easier to define the most relevant risks. It provided a clear rationale for these decisions. It is unusual to use the attacker's motivation for risk assessment. However, it might be a solution for the standardized risk assessment methods that need to estimate probability and consequence. When the motivation of attackers and attacks domain were determined, the relevant risks for the project were easily defined.

\subsection{Threats to validity}

The validity assessment is an essential part of any empirical study. Validity discussions typically involve Construct Validity, Internal Validity, External Validity, and Conclusion Validity. This study includes some threats to validity and is analyzed following a checklist by Robson~\cite{robson2002real}, which consists of concrete approaches for enhancing the validity of case studies.

\begin{table*}[htbp]
\centering
\caption[]{summarizes all findings through the case study research and compares the selected three case studies against each other.}

 \includegraphics[trim=115 50 115 50,clip,width=1\textwidth]{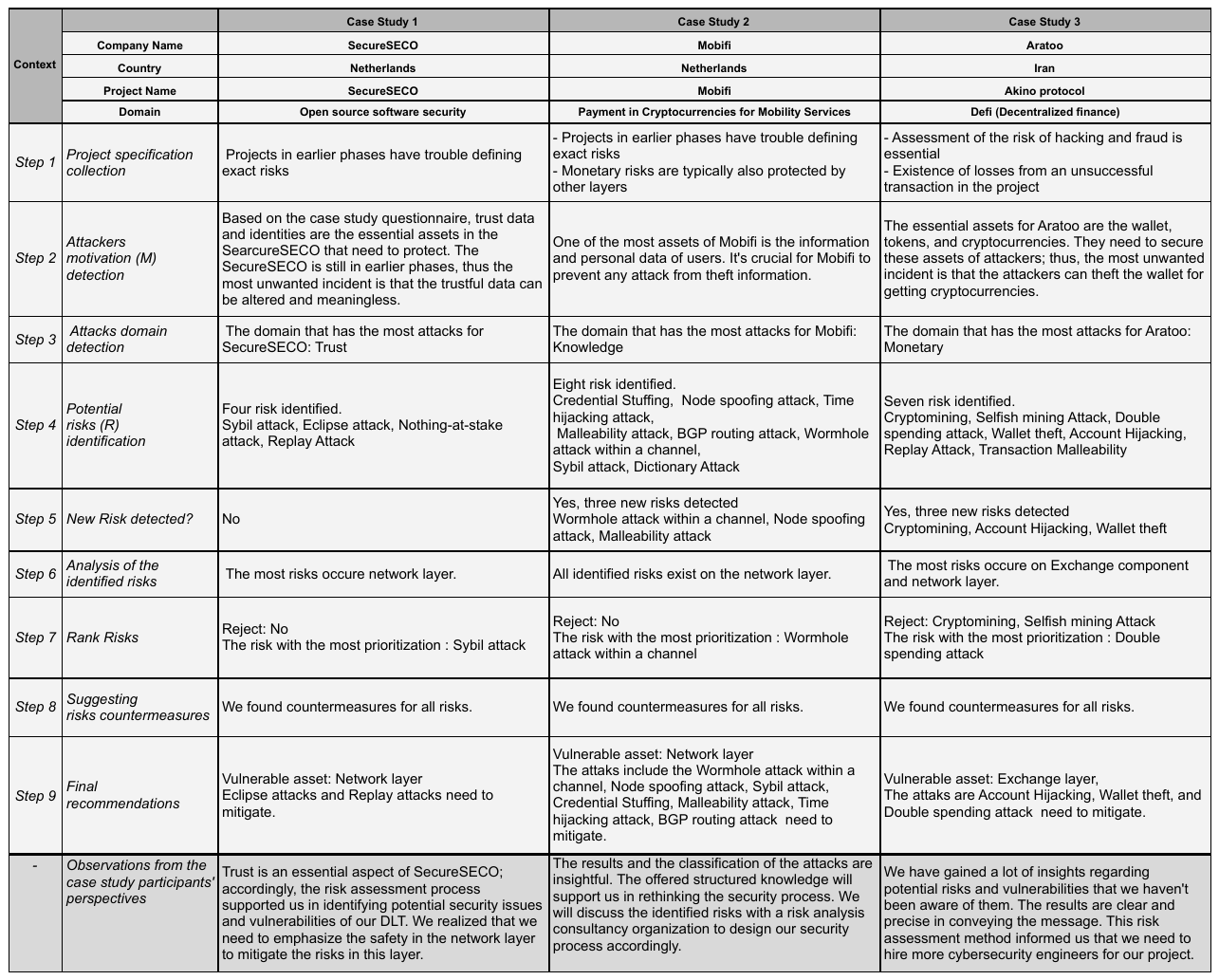}
\centering
\label{fig:Comparison}
\end{table*}

\noindent\textbf{Construct validity} refers to whether an accurate operational measure or test has been used for the concepts being studied. In literature, SRAMDA is defined as a method for assessing the risks of DLT-based applications, which contains nine steps. We followed the SRAMDA method and the nine-step risk assessment to mitigate the threats to construct validity. Moreover, we employed document analysis and literature study as two knowledge acquisition techniques to capture knowledge regarding risks and threats in DLTs. The SRAMDA method has been evaluated through three real-world case studies at three different real-world enterprises in the Netherlands and Iran.

\noindent\textbf{Internal validity} attempts to verify claims about the cause-effect relationships within the context of a study. In other words, it determines whether the study is sound or not. To mitigate the threats to the method's internal validity, we emphasize that the case study participants' opinion as a measurement instrument is risky, as they may not have sufficient knowledge to make a valid judgment. We counter this risk by conducting more than one case study, assuming that the case study participants are handling their interests.

\noindent\textbf{External validity} concerns the domain to which the research findings can be generalized. External validity is sometimes used interchangeably with generalizability (feasibility of applying the results to other research settings). To mitigate threats to the research's external validity, we captured knowledge from different sources of knowledge without any regional limitations to define the constructs.

\noindent\textbf{Conclusion validity} verifies whether the methods of a study, such as the data collection method, can be reproduced with similar results. We captured knowledge from the sources of knowledge. The accuracy of the extracted knowledge was guaranteed through the protocols that were developed to define the knowledge extraction strategy and format (See~\ref{tab:riskdatabasetable}). A review protocol was proposed and applied by multiple research assistants, including bachelor's and master's students, to mitigate the threats to the research's conclusion validity. By following the framework and the protocols, we keep the knowledge extraction process consistent and check whether the acquired knowledge addresses the research questions. Moreover, we crosschecked the captured knowledge to assess the quality of the results, and we had at least two assistants extract data independently.

\subsection{Limitations} \label{limitations}

The literature study, especially the construction of the risk knowledge base and knowledge collection, was a very time-consuming phase of the research. Most of the risks include many synonyms and ambiguous descriptions, So summarizing and extracting valuable knowledge took a long time for the research. This issue led to a comprehensive overview of all the DLT risks. On the other hand, DLT is a continuously evolving subject, and the literature is still very immature. Thus, the risk database might not have been complete, as new risks arise every day from the most distant corners of the Internet. However, much meticulous energy has been put into creating the knowledge base, and the probability of a significant risk missing at that time is negligible.  
Furthermore, It is challenging to find an appropriate countermeasure for risks. Sometimes, we are faced with a risk in the assessment process; then, we realize there is no countermeasure for such risk. It is essential to understand that no countermeasure can eliminate risk. It is one of the challenges for risk assessment because applying countermeasures at least decreases the probability of occurrence of the risk in the project.

Security vulnerabilities arising from centralization, such as Total Value Locked (TVL) concentration, Miner Extractable Value (MEV) exploitations, and account abstraction, significantly compromise the integrity and confidentiality of blockchain systems. Concentrated TVL within specific protocols or platforms creates a single point of failure, rendering the entire system susceptible to targeted attacks and manipulations. MEV vulnerabilities enable malicious actors to engage in front-running and other manipulative tactics, exploiting the transaction ordering in decentralized environments. Additionally, the potential for account abstraction threatens the security and privacy of sensitive information stored within the network, allowing unauthorized access and control over user accounts.

Furthermore, economic security vulnerabilities embedded in the token design, notably the distinction between algorithmic and asset-backed stablecoins, underscore the intricacies associated with maintaining stable value and sustainable market equilibrium. The inherent volatility of algorithmic stablecoins poses challenges, as their value is subject to significant fluctuations, while the reliance on underlying assets in asset-backed stablecoins introduces risks related to asset liquidity and valuation dynamics. Bridging assets across distinct blockchain networks, particularly the transfer of assets from platforms with changing security guarantees, like wrapping assets from Solana onto Ethereum, exposes the transferred assets to the vulnerabilities inherent in the less secure network, potentially compromising the integrity and safety of the transferred assets \cite{qin2022quantifying}.

\section{Conclusion and future work}\label{CONCLUSION}

In this article, a method is provided that performs security risk assessment for DLT-based applications. The method was created through systematic analysis of in-depth knowledge about DLT, cyber security, and other risk assessment methods. The designed method was evaluated in three real-world case studies. Furthermore, 200+ risks for DLT-based applications were identified as part of the method and are also part of the contribution of this work. 

A security risk assessment method is considered successful when it provides new insights and a clear direction of where to focus while further developing (security aspects of) the project. The findings from the case studies are that the method does contribute new insights and that the method is deemed useful by developers of DLT-based applications.

A number of recommendations for future work are given. Firstly, future research may further validate the conclusions drawn in this study. For instance, the SRAMDA can be evaluated with expert interviews. So, future work may further develop this method and assess its effectiveness for other domains such as DAOs. Secondly, the importance of researching the cybersecurity domain of DLT is stressed in this paper and calls for more attention in research. For example, research into the security aspects of each layer of DLT can be performed for a more in-depth understanding of the risks in each layer. Thirdly, besides studying the risks, future work should also consider possible countermeasures. For instance, there still is no ideal solution to Sybil attacks. 

Summarizing, this article brings the risk assessment field forward by extensively identifying risks in DLT-based applications, which is a necessity for this fast developing field. Furthermore, the risk database can serve as a basis for future risk analysis and other kinds of research.



\section*{Appendix: Case Study Protocols}
\label{Appendix}
 This project captured knowledge regarding potential security threats for DLT based on an extensive literature study. Additionally, we have designed a framework for rapidly assessing threats software-producing organizations might face when developing a dApp. We will analyze your project and assess potential risks you might deal with during the meeting. The analysis results will be confidential, and we will only use them to evaluate the framework.
During this meeting, the first seven steps of the framework will be performed to define the context of your project. As you might already be noticed, the first step has already been undertaken. Step two indicates participants and the time of the meeting. Steps 3-7 are the core elements of the meeting. These steps are summarized as follows. The meeting will walk through these steps in the form of a semi-structured interview. The participants can be prepared for the meeting by reading the questionnaire. However, be aware that the questions might change or be formulated differently in the meeting.

\textbf{Step 1) Context Establishment}
\begin{itemize}
\item Can you introduce your company briefly?
\item What is the main goal of the project? Why is it being developed?
\item What technologies do you use? What consensus mechanisms? Smart contracts? Etc. Technical details are important here. 
\item At what stage of design/development is the project? What decisions are set, and which decisions are still to be taken?
\item Has your company ever participated in another cybersecurity exercise with another party?
\item What would you like to see achieved at the end of this assessment?
\item How do you think a security risk assessment could contribute to the organization? What is its purpose?
\end{itemize}

\textbf{Step 2) Determine the scope of the assessment }
\begin{itemize}
\item What is the scope of this security risk assessment for you? 
\\
For example, The scope of this risk assessment is to provide valuable insights into your company's cyber security, such that you can take the results of the assessment into account while further designing and developing your projects. 

\end{itemize}

\textbf{Step 3) Organizational characterization}

\begin{itemize}
\item Who is accountable for the top risks?
\item How is the project prepared to respond to extreme events? Is it prepared at all?
\item Are there any contradicting security needs with internal or external stakeholders? Such as, privacy is less important for us but important for our users. 
\item Can your company be deployed with some risks? When is your company secure (enough)?
\end{itemize}

\textbf{Step 4) Perfunctory risk identification}
\begin{itemize}
\item What assets need protection? What target do you wish to have analyzed?
\item What worries you the most concerning your assets? What are the main unwanted incidents that come to mind?
\item Are you already considering some identified risks while further designing and developing the project?
\item Are there already some risks you request to see in the security risk assessment?
\item Are there any organizational 'blind spots' that need attention? (Areas that are easily and often overlooked or forgotten within the organization)
\item Are you already taking certain countermeasures? If yes, what are the current countermeasures you are taking?
\item How are you planning on keeping the project secure?
\end{itemize}

\textbf{Step 5) Concluding on the assessment context}
\begin{itemize}
\item Do you think any crucial information is missing after this meeting? Do you have anything else to add? 
\end{itemize}


\bibliographystyle{unsrt}
\bibliography{refrences}

\end{document}